\documentclass[a4paper]{jpconf}

\usepackage{graphicx,epsfig,macros,macros_static,mybeam}

\def\mpi2{m_\pi^2}

\newcommand{\msbar}{{\overline{\rm MS}}}

\newcommand{\bflr}{\begin{flushright}}
\newcommand{\eflr}{\end{flushright}}
\newcommand{\bfll}{\begin{flushleft}}
\newcommand{\efll}{\end{flushleft}}
\newcommand{\ben}{\begin{enumerate}}
\newcommand{\een}{\end{enumerate}}
\newcommand{\beq}{\begin{equation}}
\newcommand{\eeq}{\end{equation}}
\newcommand{\beqn}{\begin{eqnarray}}
\newcommand{\eeqn}{\end{eqnarray}}
\newcommand{\nn}{\nonumber}

\begin{document}
\title{Phenomenological applications of non-perturbative heavy quark effective theory\footnote{based on a work in collaboration with M.~Della Morte, N.~Garron, R.~Sommer and part of the ALPHA Coll.}}

\author{Mauro Papinutto}

\address{CERN, Physics Departement, Theory Division, CH-1211 Geneva 23, Switzerland}

\ead{mauro.papinutto@cern.ch}

\begin{abstract}
We briefly review the strategy to perform non-perturbative heavy quark effective theory computations and we specialize to the case of the b quark mass which has recently been computed including the $1/m$ term. 
\end{abstract}

\section{Introduction}
Lattice QCD allows a first-principle study of non-perturbative properties of QCD (e.g. computation of the hadron spectrum, decay constants and other matrix elements). However many systematics have to be controlled in order to reliably compute quantities of phenomenological interests:
\ben
\item numerical computations can be performed only at finite non-zero lattice spacing. Results thus have to be extrapolated to the continuum limit;
\item the presence of an ultra-violet cut-off (the lattice spacing $a$) and of an infra-red cut-off (the volume $V$) constrains quark masses, and extrapolations are needed to reach the chiral and the heavy quark regimes;  
\item dynamical light quark effects are numerically expensive to simulate and are thus often neglected in the so-called quenched approximation. This approximation turned out empirically to be a good approximation in many cases and is useful in order to develop new methods and to pin down systematics. All the results presented in this contribution are obtained in this approximation.    
\een  

We now discuss the second point above in more detail. Pions have a Compton wavelength which is too large compared to lattices that can be presently simulated and in order to keep finite volume effects under control one has to simulate light quarks ($\rm u, d$) which are heavier than the physical ones and then extrapolate to the physical point by matching lattice results with the chiral effective theory.

On the contrary, the $B$ meson is too heavy and its Compton wavelength is too small compared to the lattice spacings that can be simulated at present (for 
large enough volumes). In order to keep lattice artifacts under control one is forced to simulate quark masses in the region of the charm quark and then extrapolate to the $\rm b$-quark mass. Another interesting possibility is to use heavy-quark effective theory (HQET) to describe the $\rm b$ quark. HQET can be formulated on the lattice as shown in~\cite{eichten-hill}. However, in order to be able to perform the continuum limit, it has to be renormalised non-perturbatively. A strategy has been devised and applied to the determination of the $\rm b$-quark mass in the static limit of HQET~\cite{mbstatic} and presented in a more general framework in~\cite{nonpertHQET}. We will briefly review this strategy and then describe its application to the case of the computation of the $\rm b$-quark mass including $1/m_{\rm b}$ corrections~\cite{mb1om}.
 
%\vspace*{-0.2cm}
\section{Non-perturbative HQET}

The HQET action on a lattice (up to terms of order $1/m^2$) reads 
\beq
  \label{e_shqet}
  \Shqet = a^4 \sum_x \{\heavyb D_0\heavy +  
           \omegaspin \underbrace{\heavyb(-\vecsigma\cdot \vecB)\heavy}_{\op{spin}} +
           \omegakin \underbrace{\heavyb(- {\frac12 \vecD^2 }) \heavy}_{\op{kin}}\}
           \,+ {\rm O}(1/m^2) 
\eeq 
where the first term is O(1)~\cite{eichten-hill} while the second and third terms are O($1/m$) ($\omegaspin$ and $\omegakin$ are formally of order $1/m$)~\footnote{for the unexplained notation we refer to~\cite{mb1om}}. The heavy quark fields are subject to the constraints $P_{+}\heavy=\heavy$, $\heavyb P_{+}=\heavyb$ with $P_{+}=\frac12(1+\gamma_0)$. In the following we will consider two different discretisations of the heavy quark action where the na\"\i ve parallel transporter in the covariant derivative $D_0$ is replaced by HYP1 and HYP2 gauge links (see \cite{DMSS} for details).

Also the composite operators have an $1/m$ expansion in the effective theory.
For example, the time component of the axial current, including $1/m$ terms, reads:

\bea
  \Ahqet(x) &=& \zahqet \lightb\gamma_0 \gamma_5\heavy +
                \cahqet \lightb\gamma_j \ola{D}_j \heavy \, + {\rm O}(1/m^2)
\eea
where $\cahqet = {\rm O}(\minv)$. 

In the path integral one keeps the O(1) term of the HQET action in the weight and expands the rest in powers of $1/m$ considering the higher order terms of the HQET action ($\opi{}{\nu}(x)={\rm O}(\minv^\nu)$) as operator insertions: 
\bea
   \langle {\cal{O}} \rangle = {\cal{Z}}^{-1} \int {\rm D} \phi \,
      \rme^{ -S_\mrm{light} - a^4 \sum_x \heavyb(x)D_0\heavy(x) } \,
        \,{\cal{O}}\,\left\{ 1 -  {a^4 \sum_x   \lnu{1}(x)  + \ldots }\right\}
\eea
In this way the effective theory is renormalisable order by order in $\minv$ (this statement is equivalent to assert the existence of the continuum limit due to universality) and results in an asymptotic expansion in $1/m$. It is important to notice that these properties are not automatic for an effective field theory: e.g. ChPT shares these properties while NRQCD does not.

It is important to mention that for HQET on the lattice, as for any other theory with a dimensionful (hard) cut-off, the possibility of mixing with operators of lower dimension implies that renormalisation has to be carried out non-perturbatively, in order to be able to perform the continuum limit.    

%\vspace*{-0.2cm}
\section{Matching between QCD and HQET}

The bare couplings of HQET ($m_{\rm bare}, \omega_{\rm kin}, \omegaspin, \cahqet, \zahqet, \dots$) are unknown parameters and have to be determined by matching the effective theory with QCD, thus allowing a transfer of predictivity from QCD to HQET. A way of determining the bare couplings of HQET in a non-perturbative way (thus assuring the existence of the continuum limit) is to perform the matching non-perturbatively. In practice this consists in properly choosing $N_{\rm HQET}$ (the number of HQET bare couplings to be determined at a certain order in $1/m$) QCD observables and to equate them to their corresponding expansion in HQET:
\beq
\Phiqcd_k = \Phihqet_k\,,\quad k=1,2,\ldots,N_{\rm HQET} 
\eeq
The bare couplings of HQET can consequently be extracted by solving the resulting system of equations. It is important to notice that this procedure requires to be able to simulate the $\rm b$-quark around its physical mass. How is this possible? The trick consists in performing the matching in a small physical volume and at small lattice spacing (using the Schr\"odinger Functional) in such a way that $m_{\rm b} a \ll 1$. At the same time, in small volume the expansion in $1/m$ will have corrections in $1/m_{\rm b}L$ and thus $L$ can not be chosen too small. We choose $L=L_1\approx 0.4$fm which allows both conditions $m_{\rm b} a \ll 1$ and $1/(m_{\rm b} L) \ll 1$ to be satisfied. 
 
However, physical observables need a large volume, such that the $B$-meson fits comfortably, say $L\approx 4L_1\approx 1.6\,\fm$. The connection between these different volumes is achieved by a recursive finite size method based on the concept of step scaling functions (SSFs)~\cite{LWW}:
\beq 
\Phi_k^\mrm{HQET}(2L) = F_k\left(\left\{\Phi_j^\mrm{HQET}(L),j=1,\ldots,N \right\}\right)
\eeq
All these steps allow a fully non-perturbative formulation of HQET where the continuum limit can be taken in all steps~\cite{nonpertHQET}.
 
\section{$M_\beauty$ static (at order $\minv^0$)}

We present the strategy described above in the simple case of 
$M_\beauty$ in the static approximation (i.e. at order $\minv^0$).
We start from the definition of the finite volume $B$-meson ``mass'' 
$\meff= -\partial_0 \log[\fa(x_0)]_{x_0=L/2, T=L}$ which is defined in term of the Schr\"odinger functional correlator $\fa(x_0)$ of the temporal component of the axial current (inserted at time $x_0$) with a pseudoscalar source on one boundary (the temporal extension $T$ of the Schr\"odinger functional is here chosen to be equal to the spatial size $L$). From $\Gamma$ we construct the observable $\Phi_2(L,M)^{\rm QCD}=L\Gamma(L,M)$. In the effective theory this observable has the expansion $\Phi_2(L,M)^{\rm HQET}=L[\Gamma^{\rm stat}(L) + \mhbare]$, where $\mhbare$ is the overall energy shift between the effective theory and QCD.
The matching condition $\Phi_2(L,M)^{\rm QCD}=\Phi_2(L,M)^{\rm HQET}$ in infinite volume becomes $\mB = \Estat + \mhbare$ where $\Estat=\lim_{L\rightarrow\infty}\meffstat(L)$. Multiplying by  $L_2 = 2L_1$ and eliminating $\mhbare$ one obtains:
\bea
\label{mbstat}
m_{\rm B} && =  L_2 \Estat - L_2 \meffstat(L_1) + {L_2 \over L_1} \Phihqet_2(L_1,\Mb)\nn\\[1ex] 
 && =  L_2 \Estat - L_2 \meffstat(L_1) + {L_2 \over L_1} \Phiqcd_2(L_1,\Mb)\nn\\[1ex]
 && =  L_2 \Estat - L_2 \meffstat(L_2) + 
\underbrace{L_2 \meffstat(L_2) - L_2 \meffstat(L_1)}_{= \sigmam(\gbar^2(L_1))} +
 {L_2 \over L_1} \Phiqcd_2(L_1,\Mb)\nn\\ [-2ex]
 &&  = \underbrace{L_2[\Estat - \meffstat(L_2)]}_{a\to0\, {\rm in\, HQET}}\quad +
         \qquad\quad\underbrace{\sigmam(u_1)
         }_{a\to0\, {\rm in\, HQET}}\qquad\quad
        + \quad       2 \underbrace{\overbrace{L_1\meff(L_1,{\Mb})}^{\equiv \Phi_2^{\rm QCD}(L_1,M_\beauty)}}_{a\to0\, {\rm for}\, \Mb L_1 \gg1}
\eea
where $\sigmam(\gbar^2(L_1)$ is the SSF for the static effective mass $\meffstat$. The whole procedure is represented by the following diagram: 
%%%%%%%%%%%%%%%%%%%%%%%%%%%%%%%%%%%%%%%%%%%%%%%%%%%%%%%%%%%%%%%%%%%%
%

\begin{figure}[!h]
\vspace{1.5cm}
\begin{picture}(8,20)(-45,50)
\small
%\hspace{-1.5cm}
  \unitlength 0.6cm
  \put(2,6){\ftext{experiment}}            \put(16.5,6){\ftext{Lattice with 
$a\mq\ll 1$}} 
  \put(2,4){ $\mB=5.4\,\GeV$}    \put(16,4){ $\meff(L_1,M)$} 
  \linethickness{0.3mm}\put(5.3,3.5){\vector(0,-1){1.5}}
  \linethickness{0.3mm}\put(17,3.5){\vector(0,-1){1.5}}
  \linethickness{0.3mm}
  \put(4,0.5){ $\meffstat(L_2)$}
  % \put(10,0.5){ $\meffstat(L_1)$}
  \put(16,0.5){ $\meffstat(L_1)$}
  \put(15.7,0.7){\vector(-1,0){6.0}}
  %\put( 9.7,0.7){\vector(-1,0){2.0}}
  % \put(13.5,-0.1){$\sigmam(u_0)$}
  \put(10.5,-0.1){$\sigmam(u_1)$}
  \put(10,3.0){\small $L_2 = 2 L_1$}
  \put(10,2.0){\small $u_i = \gbar^2(L_i)$}
\end{picture}
\end{figure}
\vspace{1.3cm}
%%%%%%%%%%%%%%%%%%%%%%%%%%%%%%%%%%%%%%%%%%%%%%%%%%%%%%%%%%%%%%%%%%%%

%%% Local Variables: 
%%% mode: latex
%%% TeX-master: "cag04"
%%% End: 

After having taken the continuum limit in the various parts of the last line of Eq.~\ref{mbstat} one has to solve it for $M_\beauty$ (the RGI $\beauty$-quark mass). In the simulations we fix $L_1\approx0.4\,\fm$ (from $\gbar^2(L_1)\approx3.48$), we set the light quark mass to zero and we fix the RGI quark masses of the heavy 
quark around $\Mbeauty$. In infinite volume ($L_\infty=4L_1$) we set the light quark mass to the strange quark mass because as physical input we use the mass of the $B_{\rm s}$ meson.

\begin{figure}[!t]
\psfig{figure=Phi2.eps,angle=0,width=0.38\linewidth}  
\put(25,173){\psfig{figure=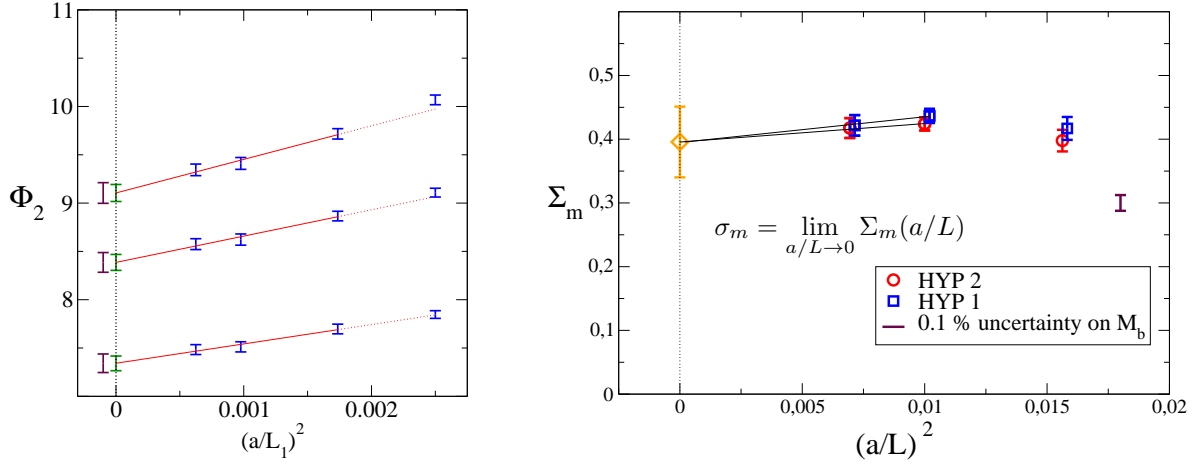,angle=-90,width=0.54\linewidth}}
\put(88,108){\onecol{3cm}{\small $$ \sigma_m=\lim_{a/L\to0} \Sigma_m(a/L)$$}}
\caption{example of continuum limit extrapolation for the QCD observable $\Phi_2^{\rm QCD}$ (for three values of the mass around $M_\beauty$) and for the static SSF $\sigma_{\rm m}.$\label{mbstatic_contlim}}
\end{figure}

\begin{figure}[!t]
\vspace*{0.3cm}
\begin{center}
\psfig{figure=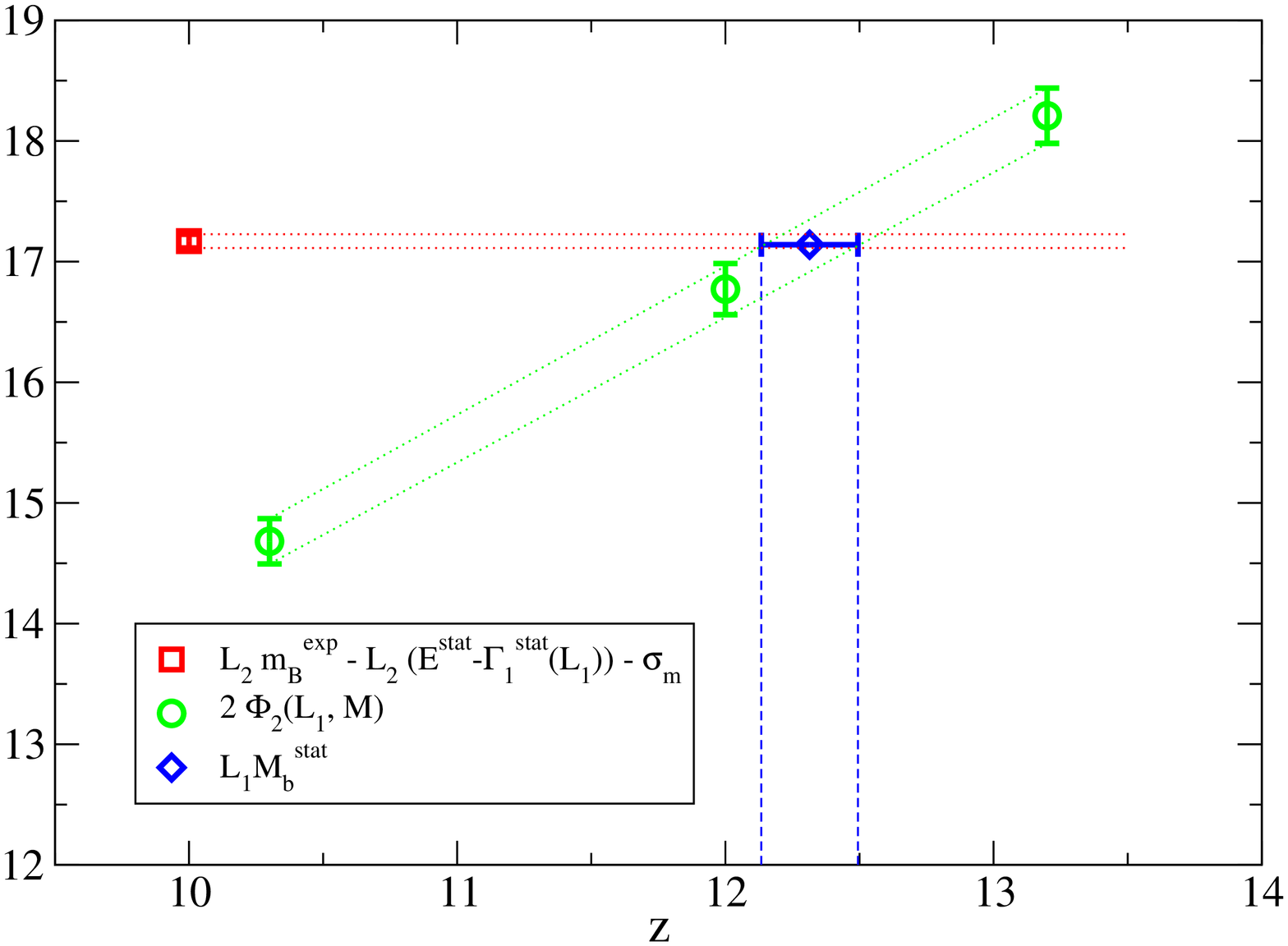,angle=-90,width=0.52\linewidth}
\end{center}
\vspace*{-0.3cm}
\caption{Graphical solution of Eq.~\ref{mbstat} to determine $M^{\rm stat}_\beauty$.\label{solution}}
\vspace*{-0.3cm}
\end{figure}

Two examples of continuum limit extrapolation are shown in Fig.~\ref{mbstatic_contlim}. The solution of Eq.~\ref{mbstat} is represented graphically in Fig.~\ref{solution}, where the red square represent the value of $L_2\mB - L_2[\Estat -\meffstat(L_1)] - \sigmam(\gbar^2(L_1))$ (with $\mB$ set to the experimental value) while the green circles correspond to the values of $2 \Phiqcd_2(L_1,M)$ for tree values of $z=L_1M$ around $L_1M_\beauty$.
From the solution of Eq.~\ref{mbstat} one obtains also the slope   
\bea 
S={1\over L_1}{\partial \over \partial M}\Phiqcd_2(L_1,M)= 0.61(5)
\eea
which will be needed in the computation of the $1/m$ corrections. The results for $M_\beauty$ in the static approximation are 
\bea
M_\beauty^{\rm stat} = 
6806 \pm 79\,{\rm MeV}
\eea
and the error is dominated by the error on the renormalisation constant of the quark mass~\cite{mb1om}.

\section{$M_\beauty$ at order $\minv$}

At order $\minv$, beyond the parameter $\mhbare$ one has to determine at least (depending on the strategy) the bare couplings $\omegakin$ and $\omegaspin$ which are of order $\minv$.
% and multiply in the HQET action the operators $\op{kin}$ and $\op{spin}$
In our case we consider the spin averaged $B$-meson mass and in this quantity $\omegaspin$ cancels out.
We rewrite Eq.~\ref{mbstat} in the case $\minv$:

\vspace*{0.3cm}
\noindent 
{\begin{tabular}{lccccccccc}
$\infty$ volume &  $\mB^{\rm av}$ & $=$ &   
$E^{\rm stat}$  & $+$ & $m_{\rm bare}$ &
$+$ &  $\omegakin E^{\rm kin}$ & &
\\
\\
Matching 1 & $\Phi_1^{\rm QCD}(L,M)$ & $=$ &
$\omegakin  R_1^{\rm kin}(L)$ & $=$ & $\Phi^{\rm HQET}_1$ & & &&
\\
\\
Matching 2 & $\meff_1(L,M)$ & $=$ & 
$\meffstat_1(L)$ & $+$ & $m_{\rm bare}$ & $+$ & $\omegakin \meffkin_1(L)$ & $=$ &
$\frac{\Phi^{\rm HQET}_2}{L}$
\end{tabular}}

\begin{figure}[!t]
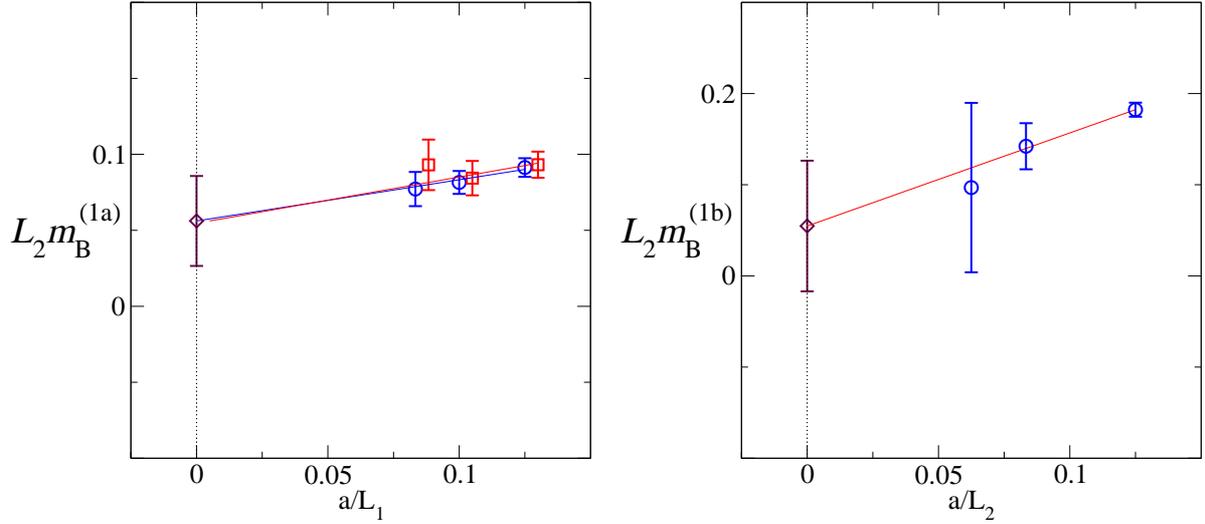

\psfig{figure=mB1a_2b2.eps,angle=0,width=0.48\linewidth}  
\put(10,0){\epsfig{figure=mB1b_2b2.eps,angle=0,width=0.48\linewidth}}
\vspace*{-0.3cm}
\caption{Continuum limit extrapolations for $\mb^{(1a)}$ (combining HYP1 and HYP2 actions) and for $\mb^{(1b)}$ (only HYP2 shown).\label{minv_contlim}}
\vspace*{-0.3cm}
\end{figure}

\vspace*{0.5cm}
\noindent 
where now $\Phi_2^{\rm QCD}\equiv L\meff_1$ and $\meff_1$ is defined similarly to $\meff$ but in terms of the Schr\"odinger functional boundary to boundary correlators (this allows a determination of $M_\beauty$ at order $\minv$ without need of determining $\cahqet$) while $\meffstat_1(L)$ and $\meffkin_1(L)$ arise naturally from the expansion of $\meff_1$ in the effective theory. $\Phi_1^{\rm QCD}(L)$ is a more complicate quantity whose definition (together that of $R_1^{\rm kin}$) can be found in~\cite{mb1om}. The important thing here is that its expansion in the effective theory is proportional to $\omegakin$, thus allowing to eliminate this parameter from the above system of equations. By eliminating also $\mhbare$ one obtains the following equation
\beq
\label{mboneoverm}
\mB^{\rm av}  = \left[ E^{\rm stat}- \meffstat_1(L)\right] +\meff_1(L,M) 
+ \left[ \frac{\Phi_1^{\rm QCD}(L,M)}{ R_1^{\rm kin}(L)}(E^{\rm kin}- \meffkin_1(L) )\right]
\eeq
In this equation we set $L=L_2=2L_1$ and we use suitable SSFs to relate $\Phi_i(L_2)$ with $\Phi_i(L_1)$:
\beq
\label{SSFs}
\Phi_1(2L,M)=\sigmakin_1(u)\Phi_1(L,M)\,,\qquad \Phi_2(2L,M)=2\Phi_2(L,M)+\sigmam(u)+\sigmakin_2(u)\Phi_1(L,M)
\eeq
The continuum SSFs are defined in terms of those at finite lattice spacing as 
$\sigma(u) = \lim_{a/L \to 0} \Sigma(u,a/L)$ where the definition of the $\Sigma$'s (neglecting their $\theta$ dependence, see~\cite{mb1om} for further details) is:
\vspace*{-0.3cm}
\bes
   \Sigma^{\rm kin}_1(u,a/L) &=& \left.{\ratonekin(2L) \over
              \ratonekin(L) }\right|_{u=\gbar^2(L)}\,, \\
   \Sigma^{\rm kin}_2(u,a/L) &=&
                \left. {2L\,[\meffkin_1(2L) - \meffkin_1(L)]
                                        \over
                                  \ratonekin(L)  }\right|_{u=\gbar^2(L)} \,, \\
   \Sigmam(u,a/L) &=& 2L
                 \left[ \meffstat_1(2L) - \meffstat_1(L)
                                \right]_{u=\gbar^2(L)} \,.
\ees
The r.h.s. of Eq.~\ref{mboneoverm} can be split in three parts $\mb = \mb^\mrm{stat}+\mb^{(1a)}+\mb^{(1b)}$. $M_\beauty^{\rm stat}$ is the solution of
$\mB^{\rm stat}(\Mb^\mrm{stat})=\mb^\mrm{av}$ while the $\minv$ correction can be found by solving the equation $\mB(\Mb^\mrm{stat}+\Mb^{(1a)}+\Mb^{(1b)})=\mb^\mrm{av}$, which translates into 
$\Mb^{(1a)}+\Mb^{(1b)}=-{1\over S}\big[\mB^\mrm{(1a)}(\Mb^{\rm stat}) + \mB^\mrm{(1b)}(\Mb^{\rm stat})\big]$. From Eqs.~\ref{mboneoverm}~and~\ref{SSFs} one obtains the formulae for $\mb^{(1a)}$, $\mb^{(1b)}$:
\bea
 \mb^{(1a)}(M) = \frac1{L_2}\sigmakin_2(u_1) \,\Phi_1(L_1,M)
        &\quad&  
\mb^{(1b)}(M) = \frac{(E^\mrm{kin} - \meffkin_1(L_2))}{R_1^\mrm{kin}}\Phi_1(L_2,M)
\eea
Some example of continuum extrapolations at order $\minv$ are shown in Fig.~\ref{minv_contlim}.
These are the most difficult steps of the computation because in both $\Sigma_2^{\rm kin}$ and $(E^\mrm{kin} - \meffkin_1(L_2))$ there are $1/a^2$ power divergences that have to cancel out and the extrapolation to the continuum limit is only linear in $a$ (in the case $(E^\mrm{stat} - \meffstat_1(L_2))$ the divergence goes only as $1/a$ and it extrapolates - as is the case for $\Phi_2^{\rm QCD}$ and $\Sigma_{\rm m}$ - quadratically in $a$). The $1/m$ correction to $\Mb$ are then 
\bea
    \Mb^{(1)} &=& \Mb^{(1a)} + \Mb^{(1b)} \nn\\
M_b^{\rm (1a)} &=& -{{\sigmakin_2(\gbar^2(L_1))\Phi_1(L_1,M_b^{\rm stat} )} \over S\,L_2} = -25(13)\,\MeV\nn\\
M_b^{\rm (1b)} &=& -\frac{(E^\mrm{kin} - \meffkin_1(L_2)) \Phi_1(L_2,M)}{S R_1^\mrm{kin}} = -24(32)\,\MeV
\eea
and in the $\msbar$ scheme:
\bea
  \mbeauty(\mbeauty) &=& \mbeauty^{\rm stat} + \mbeauty^{(1)} \nn\\
  \mbeauty^{\rm stat} &=& 4.347(48) \,\GeV \,,\quad
  \mbeauty^{(1)} = -0.027(22) \,\GeV \,.
\eea
which agrees with the value given by the Particle Data Group, despite the quenched approximation.

An alternative computation has been performed by using observables defined in terms of $\fa(x_0)$. In this case there is one additional combination of bare couplings of order $1/m$ that has to be determined and thus more step scaling functions are required. The final result is found to agree up to (small) $\rmO(1/m^2)$ corrections.
 
\vspace*{0.3cm}
We thank Rainer Sommer for a critical reading of the manuscript. 
Financial support by an EIF Marie Curie fellowship of the European 
Community's Sixth Framework Programme under contract number 
MEIF-CT-2006-040458 is acknowledged.

\end{document}